\documentclass[prd,aps,floats,preprint,preprintnumbers]{revtex4}
\usepackage{amsmath,amssymb}
\usepackage{graphicx}
\usepackage{graphics}
\usepackage{epsfig}
\usepackage{latexsym,oldgerm}
\def\beq{\begin{equation}}
\def\eeq{\end{equation}}
\def\bea{\begin{eqnarray}}
\def\eea{\end{eqnarray}}
\def\nn{\nonumber}
\def\del{\partial}
\def\ola{\overleftarrow}

\newcommand{\rma}{{\textswab{a}}}
\begin{document}
\small
\preprint{SU-4252-867 \vspace{1cm}} \setlength{\unitlength}{1mm}
\title{Direction-Dependent CMB Power Spectrum and Statistical Anisotropy from Noncommutative Geometry
\vspace{0.5cm}}
\author{E. Akofor$^{a}$}\thanks{eakofor@phy.syr.edu}\author{ A. P.
Balachandran$^{a}$}\thanks{bal@phy.syr.edu} \author{S. G.
Jo$^{a,b}$}\thanks{sgjo@knu.ac.kr}
\author{A. Joseph$^{a}$}\thanks{ajoseph@phy.syr.edu} \author{B. A.
Qureshi$^{a, c}$}\thanks{bqureshi@stp.dias.ie}
\affiliation{$^{a}$Department of Physics, Syracuse University, Syracuse, NY
13244-1130, USA \\
$^{b}$~Department of Physics, Kyungpook National University, Daegu, 702-701,
Korea\thanks{Permanent address}\\
$^{c}$~School of Theoretical Physics, Dublin Institute for Advanced Studies, 10 Burlington Road, Dublin 4, Ireland}
\begin{abstract}
\vspace{0.5cm} Modern cosmology has now emerged  as a testing ground
for theories beyond the standard model of particle physics. In this
paper, we consider quantum fluctuations of the inflaton scalar field
on certain noncommutative spacetimes and look for noncommutative
corrections in the cosmic microwave background (CMB) radiation.
Inhomogeneities in the distribution of large scale structure and
anisotropies in the CMB radiation can carry traces of
noncommutativity of the early universe. We show that its power
spectrum becomes direction-dependent when spacetime is
noncommutative. (The effects due to noncommutativity can be observed
experimentally in the distribution of large scale structure of
matter as well.) Furthermore, we have shown that the probability
distribution determining the temperature fluctuations is not
Gaussian for  noncommutative spacetimes.

\end{abstract}
\maketitle
\section{INTRODUCTION}\label{sec:intro}
The CMB radiation shows  how the universe was like when it was only
$400, 000$ years old. If photons and baryons were in equilibrium
before they decoupled from each other, then the CMB radiation we
observe today should have a black body spectrum indicating a smooth
early universe. But in 1992, the Cosmic Background Explorer (COBE)
satellite detected anisotropies in the CMB radiation, which led to
the conclusion that the early universe was not smooth: There were
small perturbations in the photon-baryon fluid.

The theory of inflation  was introduced \cite{guth, Linde, Albrecht}
to resolve the fine tuning problems associated with the standard Big
Bang cosmology. An important property of inflation is that it can
generate irregularities in the universe, which may lead to the
formation of structure. Inflation is assumed to be driven by a
classical scalar field that accelerates the observed universe
towards a perfect homogeneous state. But we live in a quantum world
where perfect homogeneity is never attained. The classical scalar
field has quantum fluctuations around it and these fluctuations act
as seeds for the primordial perturbations over the smooth universe.
Thus according to these ideas, the early universe had
inhomogeneities and we observe them today in the distribution of
large scale structure and anisotropies in the CMB radiation.

Physics at Planck  scale could be radically different. It is the
regime of string theory and quantum gravity. Inflation stretches a
region of Planck size into cosmological scales. So, at the end of
inflation, physics at Planck region should leave its signature on
the cosmological scales too.

There are indications  both from quantum gravity and string theory
that spacetime is noncommutative with a length scale of the order of
Planck length. In this paper we explore the consequences of such
noncommutativity for CMB radiation in the light of  recent
developments in the field of noncommutative quantum field theories
relating to deformed Poincar\'e symmetry.

The early universe and CMB in the noncommutative framework have been
addressed in many places \cite{Greene, Lizzi, Brandenberger1, Huang,
Brandenberger2, BalBB, Fatollahi2, Fatollahi1}. In \cite{Greene},
the noncommutative parameter $\theta_{\mu \nu} = -\theta_{\nu \mu}
=\textrm{constants}$ with $\theta_{0i} =0$, ($\mu, \nu = 0, 1, 2,
3$, with $0$ denoting time direction), characterizing the Moyal
plane is scale dependent, while \cite{Brandenberger1,
Brandenberger2, Huang} have considered noncommutativity based on
stringy space-time uncertainty relations. Our approach differs from
these authors since our quantum fields obey twisted statistics, as
implied by the deformed Poincar\'e symmetry in quantum theories.

We organize the paper as follows: In section II, we discuss how
noncommutativity breaks the usual Lorentz invariance and indicate
how this breaking can be interpreted as invariance under a  deformed
Poincar\'e symmetry. In section III, we write down an expression for
a scalar quantum field in the noncommutative framework and show how
its two-point function is modified. We review the theory of
cosmological perturbations and (direction-independent) power
spectrum for $\theta_{\mu \nu}=0$ in section IV. In section V, we
derive the power spectrum for the noncommutative Groenewold-Moyal
plane ${\cal A}_{\theta}$ and show that it is direction-dependent
and breaks statistical isotropy. In section VI, we compute the
angular correlations using this power spectrum and show that there
are nontrivial ${\cal O}( \theta^{2})$ corrections to the CMB
temperature fluctuations.  Next, in section VII, we discuss the
modifications of the $n$-point functions for any $n$ brought about
by a non-zero $\theta^{\mu \nu}$ and show in particular that the
underlying probability distribution is not Gaussian. The paper
concludes with section VIII.

\section{Noncommutative Spacetime and Deformed Poincar\'e Symmetry}
At energy scales close to the Planck scale, the quantum nature of
spacetime is expected to become important. Arguments based on
Heisenberg's uncertainty principle and Einstein's theory of
classical gravity suggest that spacetime has a noncommutative
structure at such length scales \cite{Doplicher}. We can model such
spacetime noncommutativity by the commutation relations
\cite{Connes, Madore, Landi, Bondia} \beq \label{eq:nonSpaceTime}
[\widehat{x}_{\mu}, \widehat{x}_{\nu}] = i \theta_{\mu \nu} \eeq
where $\theta_{\mu \nu} = - \theta_{\nu \mu}$ are constants and
$\widehat{x}_{\mu}$ are the coordinate functions of the chosen
coordinate system: \beq \widehat{x}_{\mu}(x) = x_{\mu}. \eeq

The above relations depend on choice of  coordinates. The
commutation relations given in eqn. (\ref{eq:nonSpaceTime}) only
hold in special coordinate systems and will look quite complicated
in other coordinate systems. Therefore,  it is important to know
that in which coordinate system the above simple form for the
commutation relations holds. For cosmological applications, it is
natural to assume that eqn. (\ref{eq:nonSpaceTime}) holds in a
comoving frame, the coordinates in which galaxies are freely
falling. Not only does it make the analysis and comparison with the
observation easier, but also make the time coordinate  the proper
time for us (neglecting the small local accelerations).

The  relations (\ref{eq:nonSpaceTime}) are  not invariant under
naive Lorentz transformations either. But they are invariant under a
deformed Lorentz Symmetry \cite{chaichian}, in which the coproduct
on the Lorentz group  is deformed while the group structure is kept
intact, as we briefly explain below.

The Lie algebra ${\cal P}$ of the Poincar\'e group has generators (basis) $M_{\alpha \beta}$ and $P_{\mu}$. The subalgebra of infinitesimal generators $P_{\mu}$ is abelian and we can make use of this fact to construct a twist element ${\cal F}_{\theta}$ of the underlying quantum group theory \cite{drinfeld, chaichian2, chari}. Using this twist element, the coproduct of the universal enveloping algebra ${\cal U}({\cal P})$ of the Poincar\'e algebra can be deformed in such a way that it is compatible with the above commutation relations.

The coproduct $\Delta_{0}$ appropriate for $\theta_{\mu \nu} =0$ is a symmetric map from ${\cal U}({\cal P})$ to ${\cal U}({\cal P}) \otimes {\cal U}({\cal P})$. It defines the action of ${\cal P}$ on the tensor product of representations. In the case of the generators $X$ of ${\cal P}$, this standard coproduct is
\beq
\Delta_{0}(X) = 1 \otimes X + X \otimes 1.
\eeq

The twist element is
\beq
{\cal F}_{\theta} = \textrm{exp}(-\frac{i}{2}\theta^{\alpha \beta}P_{\alpha} \otimes P_{\beta}), ~~~P_{\alpha} = -i \partial_{\alpha}.
\eeq
(The Minkowski metric with signature ($-, +, +, +$) is used to raise and lower the indices.)

In the presence of the twist, the coproduct $\Delta_{0}$ is modified to $\Delta_{\theta}$ where
\beq
\Delta_{\theta} = {\cal F}_{\theta}^{-1} \Delta_{0} {\cal F}_{\theta}.
\eeq

It is easy to see that the coproduct for translation generators are not deformed,
\beq
\Delta_{\theta}(P_{\alpha}) = \Delta_{0}(P_{\alpha})
\eeq
while the coproduct for Lorentz generators are deformed:
\bea
\Delta_{\theta}(M_{\mu \nu}) &=& 1 \otimes M_{\mu \nu} + M_{\mu \nu} \otimes 1 - \frac{1}{2}\Big[(P\cdot \theta)_{\mu}\otimes P_{\nu}-P_{\nu}\otimes (P\cdot \theta)_{\mu} - (\mu \leftrightarrow \nu) \Big], \nn \\
(P \cdot \theta)_{\lambda} &=& P_{\rho}\theta^{\rho}_{\lambda}.
\eea

The algebra ${\cal A}_{0}$ of functions on the Minkowski space ${\cal M}^{4}$ is commutative with the commutative multiplication $m_{0}$:
\beq
m_{0} (f \otimes g)(x) = f(x)g(x).
\eeq

The Poincar\'e algebra acts on ${\cal A}_{0}$ in a well-known way
\beq
P_{\mu}f(x) = -i \partial_{\mu}f(x), \; \; \;  M_{\mu \nu}\; f(x) = -i(x_{\mu}\partial_{\nu} - x_{\nu}\partial_{\mu})f(x).
\eeq

It acts on tensor products $f \otimes g$ using the coproduct $\Delta_{0}(X)$.

This commutative multiplication is changed in the Groenewold-Moyal algebra ${\cal A}_{\theta}$ to $m_{\theta}$:
\beq
m_{\theta} (f\otimes g)(x) = m_{0}\Big[\textrm{e}^{-\frac{i}{2}\theta^{\alpha \beta}P_{\alpha}\otimes P_{\beta}} \; f \otimes g\Big](x) = (f \star g)(x).
\eeq

Equation (\ref{eq:nonSpaceTime}) is a consequence of this $\star$-multiplication:
\bea
[\widehat{x}_{\mu}, \widehat{x}_{\nu}]_{\star} &=& m_{\theta} \; (\widehat{x}_{\mu} \otimes \widehat{x}_{\nu} - \widehat{x}_{\nu} \otimes \widehat{x}_{\mu}) = i \theta_{\mu \nu}.
\eea

The Poincar\'e algebra acts on functions $f \in {\cal A}_{\theta}$ in the usual way while it acts on tensor products $f \otimes g \in {\cal A}_{\theta} \otimes {\cal A}_{\theta}$ using the coproduct $\Delta_{\theta}(X)$ \cite{chaichian, Aschieri}.

Quantum field theories can be constructed on the noncommutative
spacetime ${\cal A}_{\theta}$ by replacing ordinary multiplication
between the fields by $\star$-multiplication and deforming
statistics as we discuss below \cite{bal, uv-ir, bal-statuv-ir,
bal-sasha-babar}. These theories are invariant under the deformed
Poincar\'e action \cite{chaichian, Aschieri, bal-statuv-ir,
bal-sasha-babar} under which $\theta_{\mu\nu}$ is invariant. It is
thus possible to observe $\theta_{\mu\nu}$ without violating
deformed Poincar\'e symmetry. But of course they are not invariant
under the standard undeformed action of the Poincar\'e group as
shown for example by the observability of $\theta_{\mu\nu}$.
\section{Quantum Fields in Noncommutative Spacetime}
It can be shown immediately that the action of the deformed coproduct is not compatible with standard statistics \cite{bal-statuv-ir}. Thus for $\theta^{\mu \nu} =0$, we have the axiom in quantum theory that the statistics operator $\tau_{0}$ defined by
\beq
\tau_{0} \; (\phi \otimes \chi) = \chi \otimes \phi
\eeq
is superselected. In particular, the Lorentz group action must and {\it does} commute with the statistics operator,
\beq
\tau_{0} \Delta_{0}(\Lambda) = \Delta_{0}(\Lambda)\tau_{0},
\eeq
where $\Lambda \in {\cal P}^{\uparrow}_{+}$, the connected component of the Poincar\'e group.

Also all the states in a given superselection sector are eigenstates of $\tau_{0}$ with the same eigenvalue. Given an element $\phi \otimes \chi$ of the tensor product, the physical Hilbert spaces can be constructed from the elements
\beq
\Big(\frac{1\pm \tau_{0}}{2}\Big) (\phi \otimes \chi).
\eeq

Now since $\tau_{0} {\cal F}_{\theta} = {\cal F}^{-1}_{\theta} \tau_{0}$, we have that
\beq
\tau_{0} \Delta_{\theta}(\Lambda) \neq \Delta_{\theta}(\Lambda) \tau_{0}
\eeq
showing that the use of the usual statistics operator is not compatible with the deformed coproduct.

But the new statistics operator
\beq
\tau_{\theta} \equiv {\cal F}^{-1}_{\theta} \tau_{0} {\cal F}_{\theta}, \; \; \; \tau_{\theta}^{2} = 1 \otimes 1
\eeq
does commute with the deformed coproduct.

The two-particle state $|p, q\rangle_{S_{\theta}, A_{\theta}}$ for bosons and fermions obeying deformed statistics is constructed as follows:
\bea
|p, q\rangle_{S_{\theta}, A_{\theta}} &=& |p\rangle \otimes_{_{S_{\theta}, A_{\theta}}} |q\rangle =\Big(\frac{1 \pm \tau_{\theta}}{2}\Big) (|p\rangle \otimes |q\rangle)\nn \\
&=& \frac{1}{2}\Big(|p\rangle \otimes |q\rangle \pm \textrm{e}^{-i p_{\mu}\theta^{\mu \nu}q_{\nu}}|q\rangle \otimes |p\rangle\Big).
\eea

Exchanging $p$ and $q$ in the above, one finds
\beq
\label{eq:pq-qp}
|p, q\rangle_{S_{\theta}, A_{\theta}} = \pm \; \textrm{e}^{-i p_{\mu}\theta^{\mu \nu}q_{\nu}}|q, p\rangle_{S_{\theta}, A_{\theta}}.
\eeq

In Fock space, the above two-particle state is constructed from a second-quantized field $\varphi_{\theta}$ according to
\bea
\frac{1}{2}\langle0|\varphi_{\theta}(x_{1})\varphi_{\theta}(x_{2}) a_{\bf q}^{\dagger}a_{\bf p}^{\dagger}|0\rangle &=& \Big(\frac{1\pm \tau_{\theta}}{2}\Big) (e_{p} \otimes e_{q})(x_{1}, x_{2})\nn \\
&=& (e_{p} \otimes_{S_{\theta}, A_{\theta}} e_{q})(x_{1}, x_{2})\nn \\
&=& \langle x_{1}, x_{2}|p, q\rangle_{S_{\theta}, A_{\theta}}
\eea
where $\varphi_{0}$ is a boson(fermion) field associated with $|p, q\rangle_{S_{0}}$ ($|p, q\rangle_{A_{0}}$).

On using eqn. (\ref{eq:pq-qp}), this leads to the commutation relation
\beq
a_{\bf p}^{\dagger}a_{\bf q}^{\dagger} =  \pm \; \textrm{e}^{i p_{\mu}\theta^{\mu \nu}q_{\nu}}\; a_{\bf q}^{\dagger}a_{\bf p}^{\dagger}.
\eeq

Let $P_{\mu}$ be the Fock space momentum operator. (It is the representation of the translation generator introduced previously. We use the same symbol for both.) Then the operators $a_{\bf p}$ , $a_{\bf p}^{\dagger}$ can be written as follows:
\beq
\label{eq:cp-ap}
a_{\bf p} = c_{\bf p} \; \textrm{e}^{-\frac{i}{2}p_{\mu}\theta^{\mu \nu}P_{\nu}}, \; \; \; a_{\bf p}^{\dagger} = c_{\bf p}^{\dagger} \; \textrm{e}^{\frac{i}{2}p_{\mu}\theta^{\mu \nu}P_{\nu}}\; ,
\eeq
$c_{\bf p}$'s being $\theta^{\mu \nu}=0$ annihilation operators.

The map from $c_{\bf p}, c_{\bf p}^{\dagger}$ to $a_{\bf p}, a_{\bf p}^{\dagger}$ in eqn. (\ref{eq:cp-ap}) is known as the ``dressing transformation" \cite{Grosse, Faddeev-Zamolodchikov}.

In the noncommutative case, a free spin-zero quantum scalar field of mass $m$ has the mode expansion
\beq
\label{eq:modeExp}
\varphi_{\theta}(x) =\int \frac{d^{3} p}{(2\pi)^{3}} \; (a_{\bf p}\; \textrm{e}_{p}(x) + a_{\bf p}^{\dagger} \; \textrm{e}_{-p}(x))
\eeq
where
\beq
\textrm{e}_{p}(x) = \textrm{e}^{-i\; p\cdot x}, \; \; p \cdot x = p_{0}x_{0} - {\bf p}\cdot {\bf x}, \; \; \; p_{0} = \sqrt{{\bf p}^{2} + m^{2}} > 0. \nn
\eeq

The deformed quantum field $\varphi_{\theta}$ differs form the undeformed quantum field $\varphi_{0}$ in two ways: $i$.) $\textrm{e}_{p}$ belongs to the noncommutative algebra of ${\cal M}^{4}$ and $ii$.) $a_{\bf p}$ is deformed by statistics. The deformed statistics can be accounted for by writing \cite{gauge-gravity}
\beq
\label{eq:twistedfield}
\varphi_{\theta} = \varphi_{0}\; \textrm{e}^{\frac{1}{2}\overleftarrow{\partial} \wedge P}
\eeq
where
\beq
\overleftarrow{\partial} \wedge P \equiv \overleftarrow{\partial}_{\mu}\theta^{\mu \nu}P_{\nu}.
\eeq

It is easy to write down the $n$-point correlation function for the deformed quantum field $\varphi_{\theta}(x)$ in terms of the undeformed field $\varphi_{0}(x)$:
\beq
\langle 0| \varphi_{\theta}(x_1) \varphi_{\theta}(x_2) \cdots \varphi_{\theta}(x_n)|0 \rangle =\langle 0|\varphi_0 (x_1) \varphi_0(x_2) \cdots \varphi_0 (x_n)|0\rangle \;
\textrm{e}^{(-{i\over2}\sum^n_{J=2}\sum^{J-1}_{I=1}\ola{\del}_{x_{I}} \wedge \ola{\del}_{x_{J}})}.
\eeq

On using
\beq
\varphi_{\theta}(x)=\varphi_{\theta}({\bf x}, t)= \int \frac{d^{3} k}{(2\pi)^{3}} \; \Phi_{\theta}({\bf k}, t) \; \textrm{e}^{i {\bf k} \cdot {\bf x}},
\eeq
we find for the vacuum expectation values, in momentum space
\bea
&&\langle 0| \Phi_{\theta}({\bf k}_1, t_1) \Phi_{\theta}({\bf k}_2, t_2) \cdots \Phi_{\theta}({\bf k}_n, t_n)|0\rangle =\textrm{e}^{({i\over2}\sum_{J>I}{\bf k}_I \wedge
{\bf k}_J)} \langle 0| \Phi_0({\bf k}_1, t_1 + {\vec{\theta}^{0} \cdot {\bf k}_2+\vec{\theta}^{0}\cdot {\bf k}_3+
\cdots+\vec{\theta}^{0}\cdot {\bf k}_n \over 2}) \times \nn \\&&\Phi_0({\bf k}_2,
t_2+{-\vec{\theta}^{0}\cdot {\bf k}_1+\vec{\theta}^{0}\cdot {\bf k}_3+\cdots+\vec{\theta}^{0}\cdot {\bf k}_n \over 2}) \cdots \Phi_0({\bf k}_n,
t_n+{-\vec{\theta}^{0}\cdot {\bf k}_1-\vec{\theta}^{0}\cdot {\bf k}_2-\cdots-\vec{\theta}^{0}\cdot {\bf k}_{n-1} \over 2})|0\rangle
\eea
where
\beq
\label{eq:vecTheta}
\vec{\theta}^{0} = (\theta^{01}, \theta^{02}, \theta^{03}).
\eeq

Since the underlying Friedmann-Lema\^itre-Robertson-Walker (FLRW) spacetime has spatial translational invariance,
\beq
{\bf k}_1 + {\bf k}_2 + \cdots + {\bf k}_n = 0,\nn
\eeq
the $n$-point correlation function in momentum space becomes
\bea
\label{eq:nthetacorr}
&&\langle 0| \Phi_{\theta}({\bf k}_1, t_1) \Phi_{\theta}({\bf k}_2, t_2) \cdots \Phi_{\theta}({\bf k}_n, t_n)|0\rangle
= \textrm{e}^{({i\over2}\sum_{J>I}{\bf k}_I \wedge
{\bf k}_J)}\langle 0| \Phi_0({\bf k}_1, t_1 - {\vec{\theta}^{0}\cdot {\bf k}_1 \over 2})\Phi_0({\bf k}_2, t_2
- \vec{\theta}^{0}\cdot {\bf k}_1-{\vec{\theta}^{0}\cdot {\bf k}_2\over2})\nn \\
&&\cdots \Phi_0({\bf k}_n, t_n - \vec{\theta}^{0}\cdot {\bf k}_1 -
\vec{\theta}^{0}\cdot {\bf k}_2 - \cdots -{\vec{\theta}^{0}\cdot {\bf k}_{n-1}}-{\vec{\theta}^{0}\cdot {\bf k}_n\over2})|0\rangle.
\eea

In particular, the two-point correlation function is
\beq
\label{eq:2thetacorr}
\langle 0| \Phi_{\theta}({\bf k}_1, t_1) \Phi_{\theta}({\bf k}_2, t_2)|0\rangle = \langle 0| \Phi_0({\bf k}_1, t_1-{\vec{\theta}^{0}\cdot {\bf k}_1 \over 2}) \Phi_0({\bf k}_2, t_2 - {\vec{\theta}^{0}\cdot {\bf k}_1 \over 2})|0\rangle,
\eeq
since it vanishes unless ${\bf k}_{1} + {\bf k}_{2} =0$ and hence $\textrm{e}^{(\frac{i}{2}\sum_{J>I}\; k_{I} \wedge k_{J})}=1$.

We emphasize that eqns. (27), (29) and (30) come from eqn. (20) which implies
eqns. (21), (23) and (25). They are exclusively due to deformed statistics. The $*$-product is still mandatory when taking products of $\varphi_{\theta}$ evaluated at the same point.

In standard Hopf algebra theory, the exchange operation is to be
performed using the $\mathcal{R}$-matrix times the flip operator
$\sigma$ \cite {mack1, mack2}. It is easy to check that $\mathcal{R}\sigma$ acts as
identity on any pair of factors in eqns. (27) and (29).

One can also explicitly show that the $n$-point functions are
invariant under the twisted Poincar\'e group while those of the
conventional theory are not. Hence the requirement of twisted
Poincar\'e invariance fixes the structure of $n$-point functions. These points are discussed further in \cite{bal-statuv-ir}.

It is interesting to note that the two-point correlation function is nonlocal in time in the noncommutative frame work. Also note the following: Assuming that $\theta^{\mu \nu}$ is non-degenerate, we can write it as
\bea
&&\theta^{\mu \nu} = \alpha \; \epsilon_{ab}\;  e^{\mu}_{a}\; e^{\nu}_{b} + \beta \; \epsilon_{ab}\; f^{\mu}_{a}\; f^{\nu}_{b},\nn \\
&&\alpha, \beta \neq 0, \; \; \epsilon_{ab} = -\epsilon_{ba},\; \; a, b = 1, 2\nn
\eea
where $e_{a}, e_{b}, f_{a}, f_{b}$ are orthonormal real vectors. Thus $\theta^{\mu \nu}$ defines two distinguished two-planes in ${\cal M}^{4}$, namely those spanned by $e_{a}$ and by $f_{a}$. For simplicity we have assumed that one of these planes contains the time direction, say $e_{1} : e^{\mu}_{1} = \delta^{\mu}_{0}$. The $\theta^{0i}$ part then can be regarded as defining a spatial direction $\vec{\theta}^{0}$ as given by eqn. (\ref{eq:vecTheta}).

We will make use of the modified two-point correlation functions given by eqn. (\ref{eq:2thetacorr}) when we define the power spectrum for inflaton field perturbations in the noncommutative frame work.

\section{Cosmological Perturbations and (Direction-Independent) Power Spectrum for $\theta^{\mu \nu}=0$}

In this section we briefly review how fluctuations in the inflaton field cause inhomogeneities in the distribution of matter and radiation following \cite{dodelson}.

The scalar field $\phi$ driving inflation can be split into a zeroth order homogeneous part and a first order perturbation:
\beq
\phi({\bf x}, t) = \phi^{(0)}(t)+\delta \phi({\bf x}, t)
\eeq

The energy-momentum tensor for $\phi$ is
\beq
{\cal T}^{\alpha}_{\; \; \; \beta} = g^{\alpha \nu}\frac{\partial \phi}{\partial x^{\nu}}\frac{\partial \phi}{\partial x^{\beta}} - g^{\alpha}_{\; \; \; \beta}\Big[\frac{1}{2}g^{\mu \nu}\frac{\partial \phi}{\partial x^{\mu}}\frac{\partial \phi}{\partial x^{\nu}} + V(\phi)\Big]
\eeq

We assume a spatially flat, homogeneous and isotropic (FLRW) background with the metric
\beq
ds^{2} = dt^{2} - \rma^{2}(t)d{\bf x}^{2}
\eeq
where $\rma$ is the cosmological scale factor, and nonvanishing $\Gamma$'s
\beq
\Gamma^{0}_{\; \; ij} = \delta_{ij}\rma^{2}H\; \; \;  \textrm{and}\; \; \;  \Gamma^{i}_{\; \; 0j}=\Gamma^{i}_{\; \; j0}=\delta^{i}_{j}H \nn
\eeq
where $H$ is the Hubble parameter.

In conformal time $\eta$ where $d \eta =\frac{dt}{\rma(t)}, -\infty < \eta < 0$, the metric becomes
\beq
ds^{2} = a^{2}(\eta)(d\eta^{2} - d{\bf x}^{2}),
\eeq
where $a$ is the cosmological scale factor now regarded as a function of conformal time. Using this metric we write the equation for the zeroth order part of $\phi$ \cite{dodelson},
\beq
\ddot{\phi}^{(0)} + 2aH \dot{\phi}^{(0)} + a^{2}V'\phi^{(0)} = 0,
\eeq
where overdots denote derivatives with respect to conformal time $\eta$ and $V'$ is the derivative of $V$ with respect to the field $\phi^{(0)}$. Notice that in conformal time $\eta$ we have $\frac{d a(\eta)}{d \eta} = a^{2}(\eta) H$ while in cosmic time $t$ we have $\frac{d \rma(t)}{d t}= \rma H$.

The equation for $\delta \phi$ can be obtained from the first order perturbation of the energy-momentum tensor conservation equation:
\beq
{\cal T}^{\mu}_{\; \; \; \nu;\; \mu} = \frac{\partial {\cal T}^{\mu}_{\; \; \; \nu}}{\partial x^{\mu}} + \Gamma^{\mu}_{\; \; \; \alpha \mu}{\cal T}^{\alpha}_{\; \; \; \nu} - \Gamma^{\alpha}_{\; \; \; \nu \mu}{\cal T}^{\mu}_{\; \; \; \alpha} =0.
\eeq

The perturbed part of the energy-momentum tensor $\delta T^{\mu}_{\; \; \; \nu}$ satisfies the following conservation equation in momentum space \cite{dodelson}:
\beq
\frac{\partial \delta T^{0}_{\; \; \; 0}}{\partial t} + ik_{i}\delta T^{i}_{\; \; \; 0} +3H \delta T^{0}_{\; \; \; 0}-H \delta T^{i}_{\; \; \; i} =0,
\eeq
where
\beq
T^{\mu \nu}({\bf k}, t)= \int d^{3}x~{\cal T}^{\mu \nu}({\bf x}, t)~\textrm{e}^{-i{\bf k}\cdot {\bf x}}.
\eeq

Let $\phi({\bf x}, t) = \int \frac{d^{3}k}{(2\pi)^{3}}~\tilde{\phi}({\bf k}, t)~\textrm{e}^{i{\bf k}\cdot {\bf x}}$. Writing down the perturbations to the energy-momentum tensor in terms of $\tilde{\phi}({\bf k}, t)$,
\bea
\delta T^{i}_{\; \; \; 0} &=& \frac{ik_{i}}{a^{3}} \dot{\tilde{\phi}}^{(0)} \delta \tilde{\phi} , \nn \\
\delta T^{0}_{\; \; \; 0} &=& \frac{-\dot{\tilde{\phi}}^{(0)} \dot{\delta \tilde{\phi}}}{a^{2}} - V'(\tilde{\phi}^{(0)}) \delta \tilde{\phi} , \nn \\
\delta T^{i}_{\; \; \; j} &=& \delta_{ij}\Big(\frac{\dot{\tilde{\phi}}^{(0)} \dot{\delta \tilde{\phi}}}{a^{2}} - V'(\tilde{\phi}^{(0)}) \delta \tilde{\phi} \Big) ,\nn
\eea
the conservation equation becomes
\beq
\ddot{\delta \tilde{\phi}} + 2aH \dot{\delta \tilde{\phi}} + k^{2} \delta \tilde{\phi} = 0.
\eeq

Eliminating the middle Hubble damping term by a change of variable $\zeta ({\bf k}, \eta) = a(\eta) \delta \tilde{\phi}({\bf k}, \eta)$, the above equation becomes
\beq
\label{eq:harmonic}
\ddot{\zeta}({\bf k}, \eta)+ \omega_{k}^{2}(\eta) \zeta({\bf k}, \eta) = 0, \; \; \; \omega_{k}^{2}(\eta) \equiv \Big(k^{2} -\frac{\ddot{a}(\eta)}{a(\eta)}\Big).
\eeq

The mode functions $u$ associated with the quantum operator $\hat{\zeta}$ satisfy
\beq
\label{eq:ukEqn}
\ddot{u}({\bf k}, \eta)+\Big(k^{2} -\frac{\ddot{a}(\eta)}{a(\eta)}\Big)u({\bf k}, \eta)=0
\eeq
with the initial conditions $u({\bf k}, \eta_{i}) = \frac{1}{\sqrt{2\omega_{k}(\eta_{i})}}$ and $\dot{u}({\bf k}, \eta_{i})=i \sqrt{\omega_{k}(\eta_{i})}$. Notice that these initial conditions have meaning only when $\omega_{k}(\eta_{i}) > 0$.

We can immediately write down the quantum operator associated with the variable $\zeta$,
\beq
\hat{\zeta}({\bf k}, \eta) = u({\bf k}, \eta)\hat{a}_{\bf k}+u^{*}({\bf k}, \eta)\hat{a}^{\dagger}_{\bf k},
\eeq
with the bosonic commutation relations $[\hat{a}_{\bf k}, \hat{a}_{{\bf k}'}] = [\hat{a}_{\bf k}^{\dagger}, \hat{a}_{{\bf k}'}^{\dagger}]=0$ and $[\hat{a}_{\bf k}, \hat{a}_{{\bf k}'}^{\dagger}] = (2 \pi)^{3} \delta^{3}({\bf k} - {\bf k}')$.

During inflation we have scale factor $a(\eta) \simeq -(\eta H)^{-1}$. Thus eqn. (\ref{eq:ukEqn}) takes the form \cite{dodelson}
\beq
\label{eq:evolution}
\ddot{u}+\Big(k^{2} -\frac{2}{\eta^{2}}\Big)u=0.
\eeq

When the perturbation modes are well within the horizon, $k|\eta| \gg 1$, one can obtain a properly normalized solution $u({\bf k}, \eta)$ from the conditions imposed on it at very early times during inflation. Such a solution is \cite{dodelson, Mukhanov}
\beq
\label{eq:properSol}
u({\bf k}, \eta)=\frac{1}{\sqrt{2k}} \Big(1-\frac{i}{k \eta}\Big)~\textrm{e}^{-ik(\eta-\eta_{i})}.
\eeq

The variances involving $\hat{\zeta}$ and $\hat{\zeta}^{\dagger}$ are
\bea
\label{eq:zeta-corr}
\langle 0|\hat{\zeta}({\bf k}, \eta) \hat{\zeta}({\bf k}', \eta)|0\rangle &=& 0, \nn \\
\langle 0|\hat{\zeta}^{\dagger}({\bf k}, \eta) \hat{\zeta}^{\dagger}({\bf k}', \eta)|0\rangle &=& 0, \nn \\
\langle 0|\hat{\zeta}^{\dagger}({\bf k}, \eta) \hat{\zeta}({\bf k}', \eta)|0\rangle &=& (2\pi)^{3} |u({\bf k}, \eta)|^{2} \delta^{3}({\bf k} - {\bf k}')\nn \\
&\equiv& (2\pi)^{3} P_{\zeta}({\bf k}, \eta) \delta^{3}({\bf k} - {\bf k}')
\eea
where $P_{\zeta}$ is the power spectrum of $\hat{\zeta}$. Eqn. (\ref{eq:zeta-corr}) can be treated as a general definition of power spectrum.

In the case when spacetime is commutative ($\theta^{\mu \nu} =0$), the power spectrum in eqn. (\ref{eq:zeta-corr}) is
\beq
\label{eq:zeta-corr-commu}
\langle 0|\hat{\zeta}^{\dagger}({\bf k}, \eta) \hat{\zeta}({\bf k}', \eta)|0\rangle = (2\pi)^{3} P_{\zeta}(k, \eta) \delta^{3}({\bf k} - {\bf k}').
\eeq

The Dirac delta function in eqns. (\ref{eq:zeta-corr}) and (\ref{eq:zeta-corr-commu}) shows that perturbations with different wave numbers are uncoupled as a consequence of the translational invariance of the underlying spacetime. Rotational invariance of the underlying (commutative) spacetime constraints the power spectrum $P_{\zeta}(k, \eta)$ to depend only on the magnitude of ${\bf k}$.

Towards the end of inflation, $k|\eta |$ ($-\infty < \eta <0$) becomes very small. In that case the small argument limit of eqn. (\ref{eq:properSol}),
\beq
\lim_{k|\eta| \rightarrow 0}~~u({\bf k}, \eta)=\frac{1}{\sqrt{2k}}~\frac{-i}{k \eta}~\textrm{e}^{-ik(\eta-\eta_{i})},
\eeq
gives the power spectrum $P_{\zeta}(k, \eta) =|u({\bf k}, \eta)|^{2}$. On using $\zeta({\bf k}, \eta)=a(\eta)\delta\tilde{\phi}({\bf k}, \eta)$, we write the power spectrum $P_{\delta \tilde{\phi}}$ for the scalar field perturbations \cite{dodelson}:
\beq
\label{eq:metric-power}
P_{\delta \tilde{\phi}}(k, \eta) =\frac{|u({\bf k}, \eta)|^{2}}{a(\eta)^{2}} =\frac{1}{2k^{3}}\frac{1}{a(\eta)^{2}\eta^{2}}.
\eeq
In terms of the Hubble parameter $H$ during inflation ($H \simeq -\frac{1}{a(\eta) \eta}$), the power spectrum becomes
\beq
\label{eq:powerDeltaPhi}
P_{\delta \tilde{\phi}}(k, \eta) = \frac{1}{2k^{3}} H^{2}.
\eeq

We are interested in the post-inflation power spectrum for the scalar metric perturbations since they couple to matter and radiation and give rise to inhomogeneities and anisotropies in their respective distributions which we observe. This spectrum comes from the inflaton field since the inflaton field perturbations get transferred to the scalar part of the metric.

We write the perturbed metric in the longitudinal gauge \cite{mukhanov},
\beq
ds^{2} = a^{2}(\eta)\Big[(1+2 \chi({\bf x}, \eta)) d\eta^{2} - (1-2\Psi({\bf x}, \eta))\gamma^{ij}({\bf x}, \eta)dx_{i}dx_{j}\Big],
\eeq
where $\chi$ and $\Psi$ are two physical metric degrees of freedom describing the scalar metric perturbations and $\gamma^{ij}$ is the metric of the unperturbed spatial hypersurfaces.

In our model, as in the case of most simple cosmological models, in the absence of anisotropic stress ($\delta T^{i}_{\; \; j} = 0$ for $i \neq j$), the two scalar metric degrees of freedom $\chi$ and $\Psi$ coincide upto a sign:
\beq
\label{eq:psi-phi}
\Psi = -\chi.
\eeq

The remaining metric perturbation $\Psi$ can be expressed in terms of the inflaton field fluctuation $\delta \tilde{\phi}$ at horizon crossing \cite{dodelson},
\beq
\tilde{\Psi} \Big|_{\textrm{\tiny{post inflation}}} = \frac{2}{3}aH \frac{\delta \tilde{\phi}}{\dot{\tilde{\phi}}^{(0)}}\Big|_{\textrm{\tiny{horizon crossing}}}
\eeq
where $\tilde{\Psi}$ is the Fourier coefficient of $\Psi$.

On using the general definition of power spectrum as in eqn. (\ref{eq:zeta-corr-commu}), the power spectra for $P_{\tilde{\Psi}}$ and $P_{\delta \tilde{\phi}}$ can be connected when a mode $k$ crosses the horizon, i.e. when $a(\eta)H =k$, say for $\eta = \eta_{0}$:
\beq
\label{PhiHorCross}
P_{\tilde{\Psi}}({\bf k}, \eta) = \frac{4}{9} \Big(\frac{a(\eta)H}{\dot{\tilde{\phi}}^{(0)}}\Big)^{2} P_{\delta \tilde{\phi}}\Big|_{a(\eta_{0})H=k}.
\eeq

From eqn. (\ref{eq:powerDeltaPhi}), eqn. (\ref{eq:psi-phi}) and
using \beq \label{eq:conversion}
{aH}/{\dot{\tilde{\phi}}^{(0)}}=\sqrt{4 \pi G / \epsilon} \eeq at
horizon crossing, where $G$ is Newton's gravitational constant and
$\epsilon$ is the slow-roll parameter in the single field inflation
model \cite{dodelson}, we have the power spectrum (defined as in
eqn. (\ref{eq:zeta-corr-commu})) for the scalar metric perturbation
at horizon crossing, \beq \label{eq:PpsiPphi} P_{\tilde{\Psi}}(k,
\eta(t)) = P_{\Phi_{0}}(k, \eta(t)) = \frac{16 \pi G}{9 \epsilon}
\frac{H^{2}}{2k^{3}}\Big|_{a(\eta_{0})H =k}, \eeq Here we wrote
$\Phi_{0}$ for $\tilde{\chi}$.

Note that the Hubble parameter $H$ is (nearly) constant during inflation and also it is the same in conformal time $\eta$ and cosmic time $t$. Since the time dependence of the power spectrum is through the Hubble parameter in eqn. (\ref{eq:PpsiPphi}), we have
\beq
\label{eq:ConstInTime}
P_{\Phi_{0}} (k, \eta(t))= P_{\Phi_{0}} (k, t) \equiv P_{\Phi_{0}} (k) =\textrm{constant in time}.
\eeq

The power spectrum in eqn. (\ref{eq:PpsiPphi}) is for commutative spacetime and it depends on the magnitude of ${\bf k}$ and not on its direction. In the next section, we will show that the power spectrum becomes direction-dependent when we make spacetime noncommutative.

\section{Direction-Dependent Power Spectrum}
The two-point function in  noncommutative spacetime, using eqn.
(\ref{eq:2thetacorr}), takes the form \beq \label{eq:modified2pt}
\langle 0| \Phi_{\theta}({\bf k}, \eta) \Phi_{\theta}({\bf k}',
\eta)|0\rangle = \langle 0| \Phi_0({\bf k}, \eta^{-}) \Phi_0({\bf
k}', \eta^{-})|0\rangle~, \eeq where $\eta^{-} =\eta( t -
{\vec{\theta}^{0}\cdot {\bf k} \over 2})$.

In the commutative case, the reality  of the two-point correlation
function (since the density fields $\Phi_{0}$ are real) is obtained
by imposing the condition \beq \langle \Phi_{0}({\bf k},
\eta)\Phi_{0}({\bf k}', \eta)\rangle{}^{*} = \langle \Phi_{0}(-{\bf
k}, \eta)\Phi_{0}(-{\bf k}', \eta)\rangle. \eeq

But this condition is not correct  when the fields are deformed.
That is because even if $\Phi_{\theta}$ is self-adjoint,
$\Phi_{\theta}({\bf x}, t) \Phi_{\theta}({\bf x}', t') \neq
\Phi_{\theta}({\bf x}', t') \Phi_{\theta}({\bf x}, t)$ for
space-like separations. A simple and natural modification (denoted
by subscript $M$) of the correlation function that ensures reality
involves ``symmetrization" of the product of $\varphi_{\theta}$'s or
keeping its self-adjoint part. That involves replacing the product
of $\phi_\theta$'s by half its anti-commutator,
 \beq
\frac{1}{2}[\varphi_{\theta}({\bf x}, \eta), \varphi_{\theta}({\bf
y}, \eta)]_{+} = \frac{1}{2}\Big(\varphi_{\theta}({\bf x}, \eta)
\varphi_{\theta}({\bf y}, \eta)+ \varphi_{\theta}({\bf y}, \eta)
\varphi_{\theta}({\bf x}, \eta)\Big) .\eeq (We emphasize that this
procedure for ensuring reality is a matter of choice)

 For the Fourier
modes $\Phi_\theta$, this procedure gives : \beq
\label{eq:modified2pt2} \langle \Phi_{\theta}({\bf k},
\eta)\Phi_{\theta}({\bf k}', \eta)\rangle_{M} =
\frac{1}{2}\Big(\langle \Phi_{\theta}({\bf k},
\eta)\Phi_{\theta}({\bf k}', \eta)\rangle + \langle
\Phi_{\theta}(-{\bf k}, \eta)\Phi_{\theta}(-{\bf k}',
\eta)\rangle{}^{*}\Big) \eeq

After the modification of the correlation function, the power spectrum for scalar metric perturbation takes the form
\beq
\langle \Phi_{\theta}({\bf k}, \eta)\Phi_{\theta}({\bf k}', \eta)\rangle_{M} = (2\pi)^{3} P_{\Phi_{\theta}}({\bf k}, \eta) \delta^{3} ({\bf k}+{\bf k}').
\eeq

Using eqns. (\ref{eq:metric-power}), (\ref{PhiHorCross}), (\ref{eq:modified2pt}) and (\ref{eq:modified2pt2}) we write down the modified power spectrum:
\beq
\label{eq:noncommu-power-spectrum}
P_{\Phi_{\theta}}({\bf k}, \eta) = \frac{1}{2}\Big[\frac{4}{9}\Big(\frac{a(\eta)H}{\dot{\tilde{\phi}}^{(0)}}\Big)^{2}\frac{1}{a(\eta)^{2}}\Big(|u({\bf k}, \eta^{-})|^{2} + |u(-{\bf k}, \eta^{+})|^{2} \Big)\Big].
\eeq
where $\eta^{\pm} =\eta( t \pm {\vec{\theta}^{0}\cdot {\bf k} \over 2})$. Notice that here the argument of the scale factor $a(\eta)$ is not shifted, since it is not deformed by noncommutativity.

It is easy to show that
\beq
\label{eq:modified-sol}
u({\bf k}, \eta^{\pm})=\frac{\textrm{e}^{-ik\eta^{\pm}}}{\sqrt{2k}} \Big(1-\frac{i}{k \eta^{\pm}}\Big)
\eeq
are also solutions of eqn. (\ref{eq:evolution}).

Thus on using eqn. (\ref{eq:conversion}) and the limit $k\eta^{\pm} \rightarrow 0$ of eqn. (\ref{eq:modified-sol}), the modified power spectrum is found to be
\bea
\label{eq:modifiedps}
P_{\Phi_{\theta}}({\bf k}, \eta) &=& \frac{1}{2}\Big[\frac{16 \pi G}{9 \epsilon}\frac{1}{a(\eta)^{2}}\Big(|u({\bf k}, \eta^{-})|^{2} + |u(-{\bf k}, \eta^{+})|^{2} \Big)\Big]\nn \\
&=& \frac{1}{2}\Big[\frac{16 \pi G}{9 \epsilon}\frac{1}{a(\eta)^{2}}\Big(\frac{1}{2k^{3}(\eta^{-})^{2}} + \frac{1}{2k^{3}(\eta^{+})^{2}} \Big)\Big]\nn \\
&=& \frac{8 \pi G}{9 \epsilon}\frac{1}{2k^{3}a(\eta)^{2}}\Big(\frac{1}{(\eta^{-})^{2}} + \frac{1}{(\eta^{+})^{2}} \Big).
\eea

Assuming that the Hubble parameter $H$ is nearly a constant during inflation, the conformal time \cite{dodelson}
\beq
\label{eq:EtaH}
\eta(t) \simeq \frac{-1}{Ha_{0}}~\textrm{e}^{-Ht}.
\eeq
gives an expression for $\eta^{\pm}$:
\beq
\label{eq:etapm}
\eta^{\pm} = \eta(t)~\textrm{e}^{\mp \frac{1}{2}H\vec{\theta}^{0}\cdot {\bf k}}.
\eeq

On using eqn. (\ref{eq:etapm}) in eqn. (\ref{eq:modifiedps}) we can easily write down an analytic expression for the modified primordial power spectrum at horizon crossing,
\beq
\label{eq:modifiedps2}
P_{\Phi_{\theta}}({\bf k}) = P_{\Phi_{0}}(k) \; \textrm{cosh}(H \vec{\theta}^{0}\cdot {\bf k})
\eeq
where $P_{\Phi_{0}}(k)$ is given by eqn. (\ref{eq:PpsiPphi}). Note that the modified power spectrum also respects the ${\bf k} \rightarrow -{\bf k}$ parity symmetry.

This power spectrum depends on both the magnitude and direction of ${\bf k}$ and clearly breaks rotational invariance. In the next section we will connect this power spectrum to the two-point temperature correlations in the sky and obtain an expression for the amount of deviation from statistical isotropy due to noncommutativity.

\section{Signature of Noncommutativity in the CMB Radiation}

We are interested in quantifying the effects of noncommutative scalar perturbations on the cosmic microwave background fluctuations. We assume homogeneity of temperature fluctuations observed in the sky. Hence it is a function of a unit vector giving the direction in the sky and can be expanded in spherical harmonics:
\beq
{\Delta T(\hat{n}) \over T} = \sum_{l m} a_{l m} Y_{l m}(\hat{n}),
\eeq
Here $\hat{n}$ is the direction of incoming photons.

The coefficients of spherical harmonics contain all the information encoded in the temperature fluctuations. For $\theta^{\mu \nu}=0$, they can be connected to the primordial scalar metric perturbations $\Phi_{0}$,
\beq
\label{eq:alm}
a_{lm} = 4 \pi (-i)^{l} \; \int \frac{d^{3}k}{(2 \pi)^{3}} \; \Delta_{l}(k) \Phi_{0}({\bf k}, \eta)Y_{lm}^{*}(\hat{k}),
\eeq
where $\Delta_{l}(k)$ are called {\it transfer functions}. They describe the evolutions of scalar metric perturbations $\Phi_{0}$ from horizon crossing epoch to a time well into the radiation dominated epoch.

The two-point temperature correlation function can be expanded in spherical harmonics:
\beq
\label{eq:TwoPZero}
\langle {\Delta T(\hat{n}) \over T} {\Delta T(\hat{n}') \over T} \rangle = \sum_{lml'm'} \langle a_{lm} a^{*}_{l'm'}\rangle Y^{*}_{lm}(\hat{n}) Y_{l'm'}(\hat{n}').
\eeq

The variance of $a_{lm}$'s is nonzero. For $\theta^{\mu \nu}=0$, we have
\beq
\langle a_{lm} a^{*}_{l'm'}\rangle = C_{l} \delta_{ll'} \delta_{mm'}.
\eeq

Using eqn. (\ref{eq:zeta-corr-commu}) and eqn. (\ref{eq:alm}), we can derive the expression for $C_{l}$'s for $\theta^{\mu \nu}=0$:
\bea
\langle a_{lm} a^{*}_{l'm'}\rangle &=& 16 \pi^{2}(-i)^{l-l'}\; \int \frac{d^{3}k}{(2\pi)^{3}}\frac{d^{3}k'}{(2\pi)^{3}} \; \Delta_{l}(k)\Delta_{l'}(k') \langle \Phi_{0}({\bf k}, \eta)\Phi^{*}_{0}({\bf k}', \eta)\rangle \; Y^{*}_{lm}(\hat{k})Y_{l'm'}(\hat{k}')\nn \\
&=& 16 \pi^{2}(-i)^{l-l'} \int \frac{d^{3}k}{(2 \pi)^{3}} \; \Delta_{l}(k)\Delta_{l'}(k) P_{\Phi_{0}}(k) \; Y^{*}_{lm}(\hat{k})Y_{l'm'}(\hat{k})\nn \\
&=& \frac{2}{\pi}\int dk \; k^{2} \; (\Delta_{l}(k))^{2} \; P_{\Phi_{0}}(k) \; \delta_{ll'}\delta_{mm'} = C_{l}\; \delta_{ll'} \delta_{mm'},
\eea
where $P_{\Phi_{0}}(k)$ is given by eqn. (\ref{eq:PpsiPphi}).

When the fields are noncommutative, the two-point temperature correlation function clearly depends on $\theta^{\mu \nu}$. We can still write the two-point temperature correlation as in eqn. (\ref{eq:TwoPZero}):
\beq
\langle {\Delta T(\hat{n}) \over T} {\Delta T(\hat{n}') \over T} \rangle_{_\theta} = \sum_{lml'm'} \langle a_{lm} a^{*}_{l'm'}\rangle_{_\theta} Y_{lm}(\hat{n}) Y^{*}_{l'm'}(\hat{n}').
\eeq

This gives
\bea
\label{ThetaAlmEqn}
&&\langle a_{lm} a^{*}_{l'm'}\rangle_{_\theta}=16 \pi^{2}(-i)^{l-l'} \int \frac{d^{3}k}{(2\pi)^{3}}\frac{d^{3}k'}{(2\pi)^{3}}\Delta_{l}(k)\Delta_{l'}(k') \langle \Phi_{\theta}({\bf k}, \eta)\Phi_{\theta}^{\dagger}({\bf k}', \eta)\rangle_{M} Y^{*}_{lm}(\hat{k})Y_{l'm'}(\hat{k}').~~~~
\eea

The two-point correlation function in eqn. (\ref{ThetaAlmEqn}) is calculated during the horizon crossing of the mode {\bf k}. Once a mode crosses the horizon, it becomes independent of time, so that we can rewrite the two-point function as
\beq
\langle \Phi_{\theta}({\bf k}, \eta)\Phi_{\theta}^{\dagger}({\bf k}', \eta)\rangle_{M} = (2\pi)^{3} P_{\Phi_{\theta}}({\bf k}) \delta^{3} ({\bf k} - {\bf k}')
\eeq
where $P_{\Phi_{\theta}}({\bf k})$ is given by eqn. (\ref{eq:modifiedps2}).

Thus we write the noncommutative angular correlation function as follows:
\bea
\label{ThetaAlm}
\langle a_{lm} a^{*}_{l'm'}\rangle_{_\theta}&=& 16 \pi^{2}(-i)^{l-l'}\int \frac{d^{3}k}{(2 \pi)^{3}} \;\Delta_{l}(k)\Delta_{l'}(k) P_{\Phi_{\theta}}({\bf k}) \; Y^{*}_{lm}(\hat{k})Y_{l'm'}(\hat{k}).
\eea

The regime in which the transfer functions act is well above the noncommutative length scale, so that it is perfectly legitimate to assume that the transfer functions are the same as in the commutative case.

Assuming that the $\vec{\theta^{0}}$ is along the $z$-axis, we have the expansion
\beq
\label{eq:ExrThetaKH}
\textrm{e}^{\pm \vec{H\theta^{0}}\cdot{\bf k}} = \sum_{l=0}^{\infty} i^{l} \sqrt{4\pi(2l+1)} j_{l}(\mp i\theta k H)Y_{l0}(\textrm{cos}\vartheta)
\eeq
where $\vec{\theta^{0}}\cdot{\bf k}=\theta k \; \textrm{cos}\vartheta$ and $j_{l}$ is the spherical Bessel function.

On using eqn. (\ref{eq:ExrThetaKH}) and the identities $j_{l}(-z) = (-1)^{l}j_{l}(z)$ and $j_{l}(iz) = i^{l}\; i_{l}(z)$, where $i_{l}$ is the modified spherical Bessel function, we can write eqn. (\ref{eq:modifiedps2}) as
\beq
\label{PowerThetaExp}
P_{\Phi_{\theta}} ({\bf k}) = P_{\Phi_{0}} (k)\sum_{l=0, \; l: \textrm{\tiny{even}}}^{\infty}\sqrt{4\pi(2l+1)}\; i_{l}(\theta k H)\; Y_{l0}(\textrm{cos}\vartheta).
\eeq

Using eqns. (\ref{ThetaAlm}) and (\ref{PowerThetaExp}),  we rewrite eqn. (\ref{ThetaAlm}) as,
\bea
\label{eq:ThetaAngular}
\langle a_{lm} a^{*}_{l'm'}\rangle_{_\theta} &=& \frac{2}{\pi}\int d k \sum_{l''=0, \; l'': \textrm{\tiny{even}}}^{\infty}  (i)^{l-l'} (-1)^{m}(2l''+1)  \; k^{2} \Delta_{l}(k)\Delta_{l'}(k) P_{\Phi_{0}} (k) i_{l''}(\theta kH)\nn \\
&& \times  \sqrt{(2l+1)(2l'+1)} \left( \begin{array}{ccc}
l & l' & l'' \\
0 & 0 & 0 \end{array} \right)\left( \begin{array}{ccc}
l & l' & l'' \\
-m & m' & 0 \end{array} \right),
\eea
the Wigner's 3-j symbols in eqn. (\ref{eq:ThetaAngular}) being related to the integrals of spherical harmonics:
\beq
\int d\Omega_{k}~Y_{l,-m}(\hat{k}) Y_{l'm'}(\hat{k}) Y_{l''0}(\hat{k}) = \sqrt{(2l+1)(2l'+1)(2l''+1)/4\pi} \left( \begin{array}{ccc}
l & l' & l'' \\
0 & 0 & 0 \end{array} \right)\left( \begin{array}{ccc}
l & l' & l'' \\
-m & m' & 0 \end{array} \right).
\eeq

We can also get a simplified form of eqn. (\ref{eq:ThetaAngular}) by expanding the modified power spectrum in eqn. (\ref{eq:modifiedps2}) in powers of $\theta$ up to the leading order:
\beq
\label{eq:ThetaExp}
P_{\Phi_{\theta}} ({\bf k}) \simeq P_{\Phi_{0}} (k) \Big[1 + \frac{H^{2}}{2} (\vec{\theta}^{0}\cdot {\bf k})^{2}\Big].
\eeq
A modified power spectrum of this form has been considered in \cite{Ackerman}, where the rotational invariance is broken by introducing a (small) nonzero vector. In our case, the vector that breaks rotational invariance is $\vec{\theta^{0}}$ and it emerges naturally in the framework of field theories on the noncommutative Groenewold-Moyal spacetime. We have also an exact expression for $P_{\Phi_{\theta}}(\bf k)$ in eqn. (\ref{eq:modifiedps2}).

Work is in progress to find a best fit for the data available and thereby to determine the length scale of noncommutativity.

The direction-dependent primordial power spectrum discussed in \cite{Ackerman} is considered in a model independent way in \cite{Kamionkowski} to compute minimum-variance estimators for the coefficients of direction-dependence. A test for the existence of a preferred direction in the primordial perturbations using full-sky CMB maps is performed in a model independent way in \cite{Christian}. Imprints of cosmic microwave background anisotropies from a non-standard spinor field driven inflation is considered in \cite{Mota1}. Anisotropic dark energy equation of state can also give rise to a preferred direction in the universe \cite{Mota2}.

\section{Non-causality and Noncommutative Fluctuations}

In the noncommutative frame work, the  expression for the two-point correlation function for the field $\varphi_\theta$ contains real and imaginary parts. We identified the real part with the observed temperature correlations which are real. This gave us the modified power spectrum 
\beq
P_{\Phi_{\theta}}({\bf k}) = P_{\Phi_{0}}(k) \; \textrm{cosh}(H
\vec{\theta}^{0}\cdot {\bf k}). \eeq

In this section we discuss the imaginary part of the  two-point
correlation function for the field $\varphi_\theta$. In position
space, the imaginary part of the two-point correlation function is
obtained from the ``anti-symmetrization" of the fields for a
space-like separation: \beq \label{re:non-causality}
\frac{1}{2}[\varphi_{\theta}({\bf x}, \eta), \varphi_{\theta}({\bf
y}, \eta)]_{-} = \frac{1}{2}\Big(\varphi_{\theta}({\bf x}, \eta)
\varphi_{\theta}({\bf y}, \eta)- \varphi_{\theta}({\bf y}, \eta)
\varphi_{\theta}({\bf x}, \eta)\Big). \eeq

The commutator of deformed fields, in general, is nonvanishing for
space-like separations. This  type of non-causality is an inherent
property of noncommutative field theories constructed on the
Groenewold-Moyal spacetime \cite{Bal-locality}.

To study this non-causality, we consider two smeared fields localized at ${\bf x}_{1}$ and ${\bf x}_{2}$. (The expression for non-causality diverges for conventional choices for $P_{\Phi_{0}}$ if we do not smear the fields. See after eqn. (\ref{non-causal}).) We write down smeared fields at ${\bf x}_{1}$ and ${\bf x}_{2}$.
\bea
&&\varphi(\alpha, {\bf x}_{1}) = \Big(\frac{\alpha}{\pi}\Big)^{3/2}\int d^{3}x~\varphi_{\theta}({\bf x})~e^{-\alpha({\bf x} - {\bf x}_{1})^{2}}, \\
&&\varphi(\alpha, {\bf x}_{2}) = \Big(\frac{\alpha}{\pi}\Big)^{3/2}\int d^{3}x~\varphi_{\theta}({\bf x})~e^{-\alpha({\bf x} - {\bf x}_{2})^{2}},
\eea
where $\alpha$ determines the amount of smearing of the fields. We have
\beq
\lim_{\alpha  \rightarrow \infty}\Big(\frac{\alpha}{\pi}\Big)^{3/2}\int d^{3}x~\varphi_{\theta}({\bf x})~e^{-\alpha({\bf x} - {\bf x}_{1})^{2}}=\varphi_{\theta}({\bf x}_{1}).
\eeq
The scale $\alpha$ can
be thought of as  the width of a wave packet which is a measure of
the size of the spacetime region over which an experiment is performed.

We can now write down the uncertainty relation for the fields $\varphi(\alpha, {\bf x}_{1})$ and $\varphi(\alpha, {\bf x}_{2})$ coming from eqn. (\ref{re:non-causality}):
\beq
\label{eq:noncausal}
\Delta \varphi(\alpha, {\bf x}_{1}) \Delta \varphi(\alpha, {\bf x}_{2}) \geq \frac{1}{2} \Big| \langle 0 |[\varphi(\alpha, {\bf x}_{1}), \varphi(\alpha, {\bf x}_{2})]|0\rangle \Big|
\eeq

{\it This equation is an expression for the violation of causality due to noncommutativity.}

Notice that, in momentum space, we can rewrite the commutator in terms of the primordial power spectrum $P_{\Phi_{0}}(k)$ at horizon crossing using the discussion following eqn. (\ref{eq:modified2pt2}):
\beq
\label{eq:commutator}
\frac{1}{2}\langle0|[\Phi_{\theta}({\bf k}, \eta), \Phi_{\theta}({\bf k}', \eta)]_{-}|0\rangle \Big|_{\textrm{horizon crossing}} = (2\pi)^{3}P_{\Phi_{0}}(k) \; \textrm{sinh}(H \vec{\theta}^{0}\cdot {\bf k})~\delta^{3}({\bf k}+{\bf k}')
\eeq

We can calculate the right hand side of eqn. (\ref{eq:noncausal})
\bea
&&\langle 0 |[\varphi(\alpha, {\bf x}_{1}), \varphi(\alpha, {\bf x}_{2})]|0\rangle =\Big(\frac{\alpha}{\pi}\Big)^{3}\int d^{3}x d^{3}y~\langle 0 |[\varphi_{\theta}({\bf x}), \varphi_{\theta}({\bf y})]|0\rangle~e^{-\alpha({\bf x} - {\bf x}_{1})^{2}}e^{-\alpha({\bf y} - {\bf x}_{2})^{2}}\nn \\
&&~~~~~~~~~~=\Big(\frac{\alpha}{\pi}\Big)^{3}\int d^{3}x d^{3}y \frac{d^{3}k}{(2\pi)^{3}} \frac{d^{3}q}{(2\pi)^{3}}~\langle 0 |[\Phi_{\theta}({\bf k}), \Phi_{\theta}({\bf q})]|0\rangle~e^{-i{\bf k}\cdot{\bf x}-i{\bf q}\cdot{\bf y}}e^{-\alpha[({\bf x} - {\bf x}_{1})^{2}+({\bf y} - {\bf x}_{2})^{2}]}\nn \\
&&~~~~~~~~~~=\frac{2}{(2\pi)^{3}}\Big(\frac{\alpha}{\pi}\Big)^{3}\int d^{3}x d^{3}y~d^{3}k d^{3}q~P_{\Phi_{0}}(k)~\textrm{sinh}(H\vec{\theta}^{0}\cdot {\bf k})~\delta^{3}({\bf k} + {\bf q})~e^{-i{\bf k}\cdot{\bf x}-i{\bf q}\cdot{\bf y}}e^{-\alpha[({\bf x} - {\bf x}_{1})^{2}+({\bf y} - {\bf x}_{2})^{2}]}\nn \\
&&~~~~~~~~~~=\frac{2}{(2\pi)^{3}}\Big(\frac{\alpha}{\pi}\Big)^{3}\int d^{3}x d^{3}y d^{3}k~P_{\Phi_{0}}(k)\; \textrm{sinh}(H\vec{\theta}^{0}\cdot {\bf k})~e^{-i{\bf k}\cdot({\bf x}-{\bf y})}e^{-\alpha[({\bf x} - {\bf x}_{1})^{2}+({\bf y} - {\bf x}_{2})^{2}]}\nn \\
&&~~~~~~~~~~=\frac{2}{(2\pi)^{3}}\Big(\frac{\alpha}{\pi}\Big)^{3}\int d^{3}k~P_{\Phi_{0}}(k)~\textrm{sinh}(H\vec{\theta}^{0}\cdot {\bf k})~\int d{\bf x} d{\bf y}e^{-i{\bf k}\cdot({\bf x}-{\bf y})}e^{-\alpha[({\bf x} - {\bf x}_{1})^{2}+({\bf y} - {\bf x}_{2})^{2}]}\nn \\
&&~~~~~~~~~~=\frac{2}{(2\pi)^{3}}\int
d^{3}k~P_{\Phi_{0}}(k)~\textrm{sinh}(H\vec{\theta}^{0}\cdot {\bf
k})~e^{-\frac{{\bf k}^{2}}{2 \alpha} -i{\bf k}\cdot({\bf x}_{1}-{\bf
x}_{2})} \label{non-causal-commu} \eea 
This gives for eqn. (\ref{eq:noncausal}),
\bea &&\Delta \varphi(\alpha, {\bf x}_{1}) \Delta
\varphi(\alpha, {\bf x}_{2}) \geq \Big|\frac{1}{(2\pi)^{3}}\int
d^{3}k~P_{\Phi_{0}}(k)~\textrm{sinh}(H\vec{\theta}^{0}\cdot {\bf
k})~e^{-\frac{{\bf k}^{2}}{2 \alpha} -i{\bf k}\cdot({\bf x}_{1}-{\bf
x}_{2})}\Big| \label{non-causal} \eea 
The right hand side of eqn.
(\ref{non-causal}) is divergent for  conventional asymptotic
behaviours of $P_{\Phi_{0}}$ (such as $P_{\Phi_{0}}$ vanishing for
large $k$ no faster than some inverse power of $k$) when $\alpha
\rightarrow \infty$ and thus the Gaussian width becomes zero. This
is the reason for introducing smeared fields. 

Notice that the amount of causality violation given in eqn. (\ref{non-causal}) is direction-dependent.

The uncertainty relation given in eqn. (\ref{non-causal}) is purely due to spacetime noncommutativity as it vanishes for the case $\theta^{\mu \nu} =0$. It is an expression of causality violation.

\section{Non-Gaussianity from noncommutativity}
In this section, we briefly explain how $n$-point correlation functions become non-Gaussian when the fields are noncommutative, assuming that they are Gaussian in their commutative limits.

Consider a noncommutative field $\varphi_{\theta}({\bf x}, t)$. Its first moment is obviously zero:
\beq
\langle \varphi_{\theta}({\bf x}, t)\rangle = \langle \varphi_{0}({\bf x}, t)\rangle =0.\nn
\eeq

The information about noncommutativity is contained in the higher moments of $\varphi_{\theta}$. We show that the $n$-point functions cannot be written as sums of products of two-point functions. That proves that the underlying probability distribution is non-Gaussian.

The $n$-point correlation function is
\beq
C_{n}(x_{1}, x_{2}, \cdots, x_{n}) = \langle \varphi_{\theta}({\bf x}_{1},
t_{1}) \cdots \varphi_{\theta}({\bf x}_{n}, t_{n}) \rangle
\eeq

Since $\varphi_{0}$ is assumed to be Gaussian and $\varphi_{\theta}$ is given in terms of $\varphi_{0}$ by eqn. (\ref{eq:twistedfield}), all the odd moments of $\varphi_{\theta}$ vanish.

But the even moments of $\varphi_{\theta}$ need not vanish and do not split into sums of products of its two-point functions in a familiar way.

Non-Gaussianity cannot be seen at the level of two-point functions. Consider the two-point function $C_{2}$. We write this in momentum space in terms of $\Phi_{0}$:
\beq
C_{2}=\langle \Phi_{\theta}({\bf k}_1,t_1)\Phi_{\theta}({\bf k}_2,t_2) \rangle = e^{-{i\over2}({\bf k}_2\wedge{\bf k}_1)} \Big\langle \Phi_0({\bf k}_1,t_1+{\vec{\theta}^{0}\cdot {\bf k}_2 \over 2})\Phi_0({\bf k}_2,t_2-{\vec{\theta}^{0}\cdot {\bf k}_1 \over 2}) \Big\rangle.
\eeq
where ${\bf k}_i\wedge{\bf k}_j \equiv k_{i}\theta^{ij}k_{j}$.

Making use of the translation invariance ${\bf k}_1+{\bf k}_2=0$, the above equation becomes
\bea
\langle \Phi_{\theta}({\bf k}_1,t_1)\Phi_{\theta}({\bf k}_2,t_2) \rangle &=& \Big\langle \Phi_0({\bf k}_1,t_1-{\vec{\theta}^{0}\cdot {\bf k}_1 \over 2})\Phi_0({\bf k}_2,t_2-\vec{\theta}^{0}\cdot {\bf k}_1-{\vec{\theta}^{0}\cdot {\bf k}_2 \over 2}) \Big\rangle.
\eea

Non-Gaussianity can be seen in all the $n$-point functions for $n \geq 4$  and even $n$. Still they can all be written in terms of correlation functions of $\Phi_{0}$. For example, let us consider the four-point function $C_{4}$:
\bea
&&C_{4}={\langle}\Phi_{\theta}({\bf k}_1,t_1)\Phi_{\theta}({\bf k}_2,t_2)\Phi_{\theta}({\bf k}_3,t_3)\Phi_{\theta}({\bf k}_4,t_4){\rangle}=e^{-{i\over2}({\bf k}_3\wedge{\bf k}_2+{\bf k}_3\wedge{\bf k}_1+{\bf k}_2\wedge{\bf k}_1)}\times \nn\\
&&~~~~\Big{\langle} \Phi_0({\bf k}_1,t_1-{\vec{\theta}^{0}\cdot {\bf k}_1 \over 2})\Phi_0({\bf k}_2, t_2-\vec{\theta}^{0}\cdot {\bf k}_1 - {\vec{\theta}^{0}\cdot {\bf k}_2 \over 2})\Phi_0({\bf k}_3, t_3 - \vec{\theta}^{0}\cdot {\bf k}_1 - \vec{\theta}^{0}\cdot {\bf k}_2 - {\vec{\theta}^{0}\cdot {\bf k}_3 \over 2})\times \nn \\
&&~~~~\Phi_0({\bf k}_4, t_4 - \vec{\theta}^{0}\cdot {\bf k}_1 - \vec{\theta}^{0}\cdot {\bf k}_2 - \vec{\theta}^{0}\cdot {\bf k}_3 - {\vec{\theta}^{0}\cdot {\bf k}_4 \over 2})\Big{\rangle}\nn
\eea
Here we have used translational invariance, which implies that ${\bf k}_1+{\bf k}_2+{\bf k}_3+{\bf k}_4=0$. Using this equation once more to eliminate ${\bf k}_{4}$, we find
\bea
&&C_{4}=e^{-{i\over2}({\bf k}_3\wedge{\bf k}_2+{\bf k}_3\wedge{\bf k}_1+{\bf
k}_2\wedge{\bf k}_1)}\Big{\langle}\Phi_0({\bf k}_1, t_1-{\vec{\theta}^{0}\cdot {\bf k}_1 \over 2})\Phi_0({\bf k}_2, t_2 - \vec{\theta}^{0}\cdot {\bf k}_1 - {\vec{\theta}^{0}\cdot {\bf k}_2 \over 2}) \times \nn \\
&&\times~\Phi_0({\bf k}_3, t_3 - \vec{\theta}^{0}\cdot {\bf k}_1 - \vec{\theta}^{0}\cdot {\bf k}_2 - {\vec{\theta}^{0}\cdot {\bf k}_3 \over 2})\Phi_0({\bf
k}_4, t_4 - {\vec{\theta}^{0}\cdot {\bf k}_1 + \vec{\theta}^{0}\cdot {\bf k}_2 + \vec{\theta}^{0}\cdot {\bf k}_3 \over 2})\Big{\rangle}\nn
\eea

Assuming Gaussianity for the field $\Phi_{0}$ and denoting $\Phi_{0}({\bf k}_{i}, t_{i})$ by $\Phi_{0}^{(i)}$, we have,
\bea
\langle \Phi^{(1)}_{0}\Phi^{(2)}_{0}\cdots \Phi^{(i)}_{0}\Phi^{(i+1)}_{0}\cdots\Phi^{(n)}_{0}\rangle &=& \langle \Phi^{(1)}_{0}\Phi^{(2)}_{0}\rangle \langle \Phi^{(3)}_{0}\Phi^{(4)}_{0}\rangle
\cdots \langle \Phi^{(i)}_{0}\Phi^{(i+1)}_{0}\rangle
\cdots \langle \Phi^{(n-1)}_{0}\Phi^{(n)}_{0}\rangle\nn \\
&&+~\textrm{permutations}~~(\textrm{for}~n~\textrm{even})
\eea
and
\beq
\langle \Phi^{(1)}_{0}\Phi^{(2)}_{0}\cdots \Phi^{(i)}_{0}\Phi^{(i+1)}_{0}\cdots\Phi^{(n)}_{0}\rangle  = 0~~(\textrm{for}~n~\textrm{odd}).
\eeq

Therefore $C_{4}$ is
\bea
&&{\langle}\Phi_{\theta}({\bf k}_1, t_1)\Phi_{\theta}({\bf k}_2, t_2)\Phi_{\theta}({\bf k}_3, t_3)\Phi_{\theta}({\bf
k}_4, t_4){\rangle}=e^{-{i\over2}({\bf k}_3\wedge{\bf k}_2+{\bf k}_3\wedge{\bf
k}_1+{\bf k}_2\wedge{\bf k}_1)}\times\nn\\
&&~~~~\Big(\Big{\langle}\Phi_0({\bf k}_1, t_1-{\vec{\theta}^{0}\cdot {\bf k}_1\over2})\Phi_0({\bf k}_2,
t_2-\vec{\theta}^{0}\cdot {\bf k}_1-{\vec{\theta}^{0}\cdot {\bf k}_2\over2})\Big{\rangle}\Big{\langle}\Phi_0({\bf k}_3, t_3-\vec{\theta}^{0}\cdot {\bf k}_1-\vec{\theta}^{0}\cdot {\bf k}_2-{\vec{\theta}^{0}\cdot {\bf k}_3\over2})\nn \\
&&~~~~\Phi_0({\bf k}_4, t_4 - {\vec{\theta}^{0}\cdot {\bf k}_1 + \vec{\theta}^{0}\cdot {\bf k}_2 + \vec{\theta}^{0}\cdot {\bf k}_3\over2})\Big{\rangle}+\Big{\langle}\Phi_0({\bf k}_1, t_1-{\vec{\theta}^{0}\cdot {\bf k}_1\over2})\Phi_0({\bf k}_3,
t_3-\vec{\theta}^{0}\cdot {\bf k}_1-\vec{\theta}^{0}\cdot {\bf k}_2-{\vec{\theta}^{0}\cdot {\bf k}_3\over2})\Big{\rangle}\nn \\
&&~~~~\Big{\langle}\Phi_0({\bf k}_2,
t_2-\vec{\theta}^{0}\cdot {\bf k}_1-{\vec{\theta}^{0}\cdot {\bf k}_2\over2})\Phi_0({\bf k}_4,
t_4-{\vec{\theta}^{0}\cdot {\bf k}_1+\vec{\theta}^{0}\cdot {\bf k}_2 + \vec{\theta}^{0}\cdot {\bf k}_3\over2})\Big{\rangle}+\Big{\langle}\Phi_0({\bf k}_1, t_1-{\vec{\theta}^{0}\cdot {\bf k}_1\over2})\nn \\
&&~~~~\Phi_0({\bf k}_4,
t_4-{\vec{\theta}^{0}\cdot {\bf k}_1+\vec{\theta}^{0}\cdot {\bf k}_2+\vec{\theta}^{0}\cdot {\bf k}_3\over2})\Big{\rangle}\Big{\langle}\Phi_0({\bf k}_2,
t_2-\vec{\theta}^{0}\cdot {\bf k}_1-{\vec{\theta}^{0}\cdot {\bf k}_2\over2})\nn \\
&&~~~~\Phi_0({\bf k}_3,
t_3-\vec{\theta}^{0}\cdot {\bf k}_1-\vec{\theta}^{0}\cdot {\bf k}_2-{\vec{\theta}^{0}\cdot {\bf k}_3\over2})\Big{\rangle}\Big).
\eea

Using spatial translational invariance for each two-point function, we have
\bea
&&{\langle}\Phi_{\theta}({\bf k}_1, t_1)\Phi_{\theta}({\bf k}_2, t_2)\Phi_{\theta}({\bf k}_3, t_3)\Phi_{\theta}({\bf k}_4, t_4){\rangle} = \Big[\Big{\langle}\Phi_0({\bf k}_1, t_1-{\vec{\theta}^{0}\cdot {\bf k}_1\over2})\Phi_0({\bf k}_2, t_2-{\vec{\theta}^{0}\cdot {\bf k}_1\over2})\Big{\rangle}\nn \\
&&~~\Big{\langle}\Phi_0({\bf k}_3, t_3-{\vec{\theta}^{0}\cdot {\bf k}_3\over2})\Phi_0({\bf k}_4, t_4-{\vec{\theta}^{0}\cdot {\bf k}_3\over2})\Big{\rangle}\Big]+ \textrm{e}^{-i{\bf k}_2\wedge{\bf k}_1}\Big[\Big{\langle}\Phi_0({\bf k}_1, t_1-{\vec{\theta}^{0}\cdot {\bf k}_1\over2})\Phi_0({\bf k}_3, t_3-\vec{\theta}^{0}\cdot {\bf k}_2-{\vec{\theta}^{0}\cdot {\bf k}_1\over2})\Big{\rangle}\nn \\
&&~~\Big{\langle}\Phi_0({\bf k}_2, t_2-\vec{\theta}^{0}\cdot {\bf k}_1-{\vec{\theta}^{0}\cdot {\bf k}_2\over2})\Phi_0({\bf k}_4, t_4-{\vec{\theta}^{0}\cdot {\bf k}_2\over2})\Big{\rangle}\Big]+\Big[\Big{\langle}\Phi_0({\bf k}_1, t_1-{\vec{\theta}^{0}\cdot {\bf k}_1\over2})\Phi_0({\bf
k}_4, t_4 - {\vec{\theta}^{0}\cdot {\bf k}_1\over2})\Big{\rangle}\nn \\
&&~~\Big{\langle}\Phi_0({\bf k}_2, t_2 - \vec{\theta}^{0}\cdot {\bf k}_1 - {\vec{\theta}^{0}\cdot {\bf k}_2\over2})\Phi_0({\bf
k}_3, t_3 - \vec{\theta}^{0}\cdot {\bf k}_1 - {\vec{\theta}^{0}\cdot {\bf k}_2\over2})\Big{\rangle}\Big].
\eea

Notice that the second term has a non-trivial phase which depends on the
spatial momenta ${\bf k}_{1}$ and ${\bf k}_{2}$ and the noncommutative parameter
$\theta$. As $C_{4}$ cannot be written as sums of products of $C_{2}$'s in a standard way, we see that the noncommutative probability distribution is non-Gaussian. Also it should be noted that we still cannot achieve Gaussianity of $n$-point functions even if we modify them by imposing the reality condition as we did for the two-point case.

Non-Gaussianity affects the CMB distribution and also the large scale structure (the large scale distribution of matter in the universe). We have not considered the latter. An upper bound to the amount of non-Gaussianity coming from noncommutativity can be set by extracting the four-point function from the data.
\section{Conclusions}
In this paper, we have shown that the introduction of spacetime noncommutativity gives rise to  nontrivial contributions to the CMB temperature fluctuations. The two-point correlation function in momentum space, called the power spectrum, becomes direction-dependent. Thus spacetime noncommutativity breaks the rotational invariance of the CMB spectrum. That is, CMB radiation becomes statistically anisotropic. This can be measured experimentally to set bounds on the noncommutative parameter. Currently, we \cite{numerical} are making numerical fits to the available CMB data to put bounds on $\theta$.

We have also shown that the probability distribution governing correlations of fields on the Groenewold-Moyal algebra ${\cal A}_{\theta}$ are non-Gaussian. This affects the correlation functions of temperature fluctuations. By measuring the amount of non-Gaussianity from the four-point correlation function data for temperature fluctuations, we can thus set further limits on $\theta$.

We have also discussed the signals of non-causality of non-commutative field theories in the temperature fluctuations of the CMB spectrum. It will be very interesting to test the data for such signals.
\section{Acknowledgements}
We gratefully acknowledge discussions with Cristian Armendariz-Picon and especially with Eric West. This work was partially supported by the US Department of Energy under grant number DE-FG02-85ER40231. The work of BQ was partially supported by IRCSET fellowship.

\bibliographystyle{apsrmp}

\end{document}